*June 8, 2015*

# SINGLE SOLENOIDAL MAGNETIC SYSTEM FOR IRON-FREE DETECTOR

Alexander Mikhailichenko, Cornell University, Ithaca, NY

*Abstract.* We consider a single solenoidal system for possible usage in iron-free detectors for future linear colliders.

## INTRODUCTION

In our recent publication we suggested a new type of iron-free magnetic system with multiple flux-return solenoids for a particle detector [1]. That was a finalization of development of Iron-free solenoidal systems originated in [2] where a single-solenoidal system was proposed. One disadvantage of a single solenoidal system is in slow decay of external field, generated by central solenoid. To screen the outer space we suggested usage of additional coils, surrounding central solenoid. In publication [2] we wrote "*One can consider the scheme with more coils, say additional Helmholtz type system of room temperature around whole detector can eliminate the mostly field around*". So in our case the system of room temperature coils at the outer radius and at the ends could restrict propagation of field in transverse direction. The concentric end-cap coils are eliminating propagation of field in a longitudinal direction already.

Proposal [1] was a natural evolution in developments of family of Iron-free detectors, represented in Fig.1 and Fig.2. Although with SC cables one can make a dipole magnets with $10T$ (even up to $20T$) fields, we restricted level of field in our considerations in [1] at $8T$ level, as the energy stored in a magnetic field of detector is much bigger, than in a dipole magnet (say, the LHC one), so this is making operational safety margins big enough if operated at reduced level; so this protects the SC coil against quench.

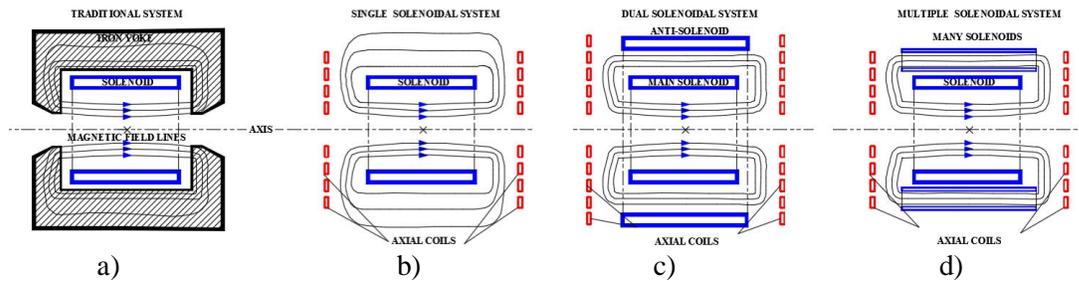

**Figure 1.** Evolution of detector magnetic system from the one with an Iron yoke a), traditional, to the Iron-free family, b)-d) (from left to right).

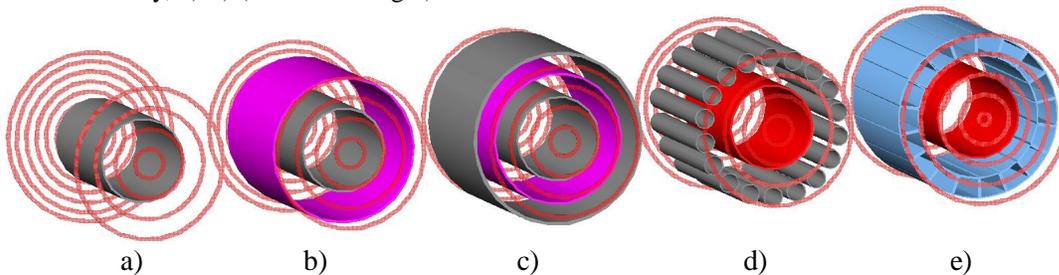

**Figure 2.** A family of Iron-free magnetic systems for detectors, 3D sketch: a)-single solenoid, b)-dual solenoids, c)-triple solenoids, d)- many return-flux solenoids. e)-many return-flux solenoids with sectorial shape [1]. Each system of solenoids surrounded by the end-cap concentric coils (front ones are shown diluted).

Dual solenoidal system, Fig.1-c), Fig.2-b) is well described in [3] as once it was a 4$^{th}$ concept detector suggested for ILC. It is interesting however to evaluate the field quality and stray fields which generated by a system with just a single solenoid, Fig.1 a), as it is the simplest one in a family of Iron-free magnetic systems for future detector.

As we are planning to continue investigation of limitations of achievable field level and its homogeneity, we first decide to clarify the option with just a single solenoid, originated in [2] for completeness.

Motivation for the Iron free detector development was clear: the Iron saturated at ~20$kG$, so further growth of field becomes reduced, effective magnetic permeability becomes equal to $\mu \sim 1$, although saturated Iron delivering its 2$T$ magnetization. With saturated Iron yoke, the homogeneity of field which was guaranteed by the Iron yoke on Fig.1-a), becomes compromised also. Also, the Iron-free detector becomes much lighter, as ~10$kT$ yoke is eliminated.

Meanwhile high magnetic field at IP and around is required by better identification of momenta of the secondary particles generated at IP. The magnetic field value together with the size of tracking system, defines momentum resolution, which is $\Delta p/p \sim p\sigma_s/(B_0 D^2)$, where $B_0$ stands for the axial field around IP, $D$ is diameter of tracking system, $\sigma_s$ is a spatial resolution of tracking system. So desire of having a high field level at IP of detector moves designers for reaching as high field as possible. A dual-solenoidal system [2] has solved this problem in part. One disadvantage of dual solenoidal system is that the outer solenoid subtracts the field from the central (inner, or main) solenoid. The level of subtraction defined by the radius of outer solenoid as namely outer solenoid squeezes the return flux between inner and outer solenoids and could reach ~2$T$.

From the other hand, absence of outer solenoid corresponds formally to the infinite enlargement of its radius. One obvious difference here is that the return field in a single solenoidal system is much smaller, than in a dual (or triple, or multiple-return) solenoidal system. So it becomes ineffective usage of this inhomogeneous field for the muon-tracking system. But if one is not interesting in a spectrometry of muons, this disadvantage might be a tolerable option. It is difficult to predict what will be interesting in far future.

So summering what was said above, it is clear that everyone wants the field level to be at IP as high as possible. The last desire supported by production of good quality SC cables. For example, new types of SC 43-strand cables described [4]-[6] allow 13$T$ field operation with 11 $kA$.

We believe that the concept of Iron-free system is an inevitable way to go for high-energy detectors in a future.

## TECHNOLOGY FOR HIGH FIELD SOLENOIDS

As we wrote above, the return field value depends on the ratio of the areas with corresponding flux. So by making the outer solenoid larger, one can reduce the field, required from outer solenoid and in reaching higher field level in the inner solenoid (less field value is subtracted).

The main solenoid made as a sectioned one with five or six enclosed solenoids feed independently. All sectioned solenoid surrounded by combined bath/indirect cooling enclosure, which represents a cold mass. Temperature of liquid Helium suggested below 1.6$^o K$, so a superfluid component is appointed for better cooling of coils. Total radial thickness of cold mass could reach 30 $cm$. Two technologies for making such solenoids considered in [1]. We prefer at the moment the one which uses the cable with slightly wedged shape embedded into a solid Al-alloy cylinder by compression and soldering[1]. So in a case of quench the current runs in a body if this cylindrical carcasses.

---

[1] Tinning Al is a known process; see for example [5]. Alloys for soldering of Al are commercially available.

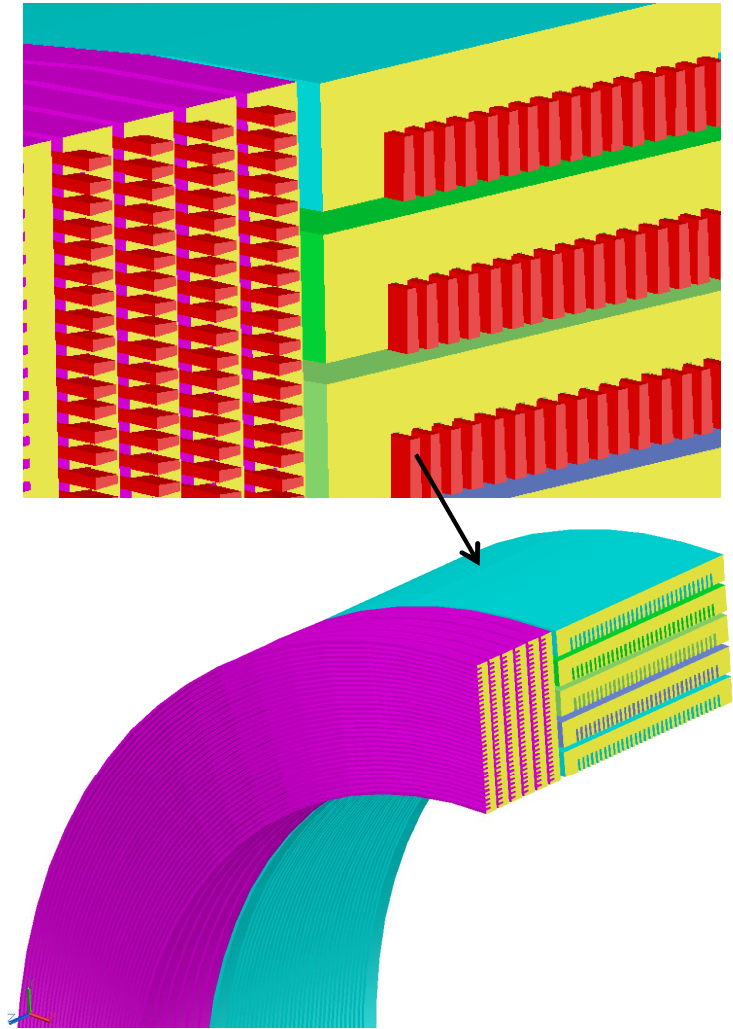

**Figure 4.** End region of central solenoid. The cables (red) compressed and soldered into Aluminum carcasses. The Helmholtz end coils inserted into flat plates with grooves, at the left. At the top-the geometry is shown zoomed.

End Helmholtz coils for this type of cabling made as a package of flat discs, where the layering in a disc is going in the radial direction, see Fig.4, left. In a regular part the grooves evenly located along the *inner* side of cylinder. Such configuration makes positioning of SC cable more rigid and magnetic pressure pushes the cable into grooves, while the cylinders accept magnetic pressure. Solenoids could be sectioned in a longitudinal (axial) direction inside a common cryostat also. Even if all stored energy (~10 *GJ*) disappeared in carcasses, temperature could be increased by ~70º C; energy evacuation system allows evacuation ~70% of stored energy, however. End cap coils are room-temperature ones with water cooled conductor.

More detailed description of design of solenoid will be done in another place, as it is common for any Iron-free detector.

General view of detector is represented in Fig.5. It looks pretty much the same as once suggested 4$^{th}$ concept detector. The space out of central solenoid is free here, see in [2]. Final focus lenses are carried by the same reinforced frame. This is done for prevention of motion of focal interaction point (IP) caused by ground motion. This is well known procedure in optics: all elements installed at the same base plate. Otherwise motion of final lens say, by 0.1 micrometer, yields movement of IP to about the same value.

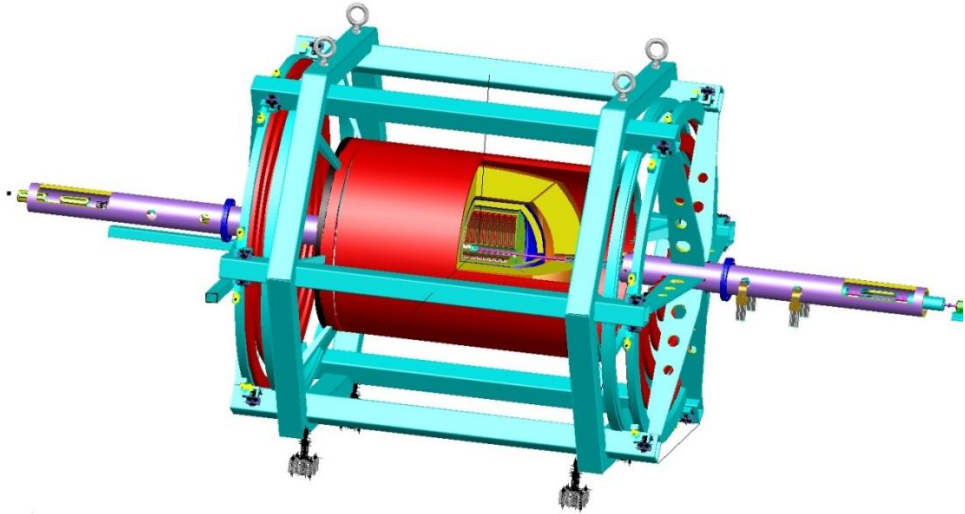

**Figure 5.** General view of the Iron-free detector with a single solenoid. By yellow painted calorimeters. *Clu-Cou* central system used for tracking [1]. FF lenses are installed on detector's frame.

## FIELDS IN A SINGLE-SOLENOIDAL SYSTEM

We believe that namely the system with multiple flux-return solenoids is a perspective one, Fig.1d), Fig2.d), e) as it delivers higher magnetic field on the axis and the less expensive processes could be used for solenoid manufacturing and handling. The system with strong return field allows making spectrometry of muons. As we have said above we will re-consider case with just a single solenoid, Fig.1 a) [3], for the completeness of description. In the following Figs. 6-13 the different graphs are describing the field distribution is this system; captions explaining each graph. Calculations done with FlexPDE.

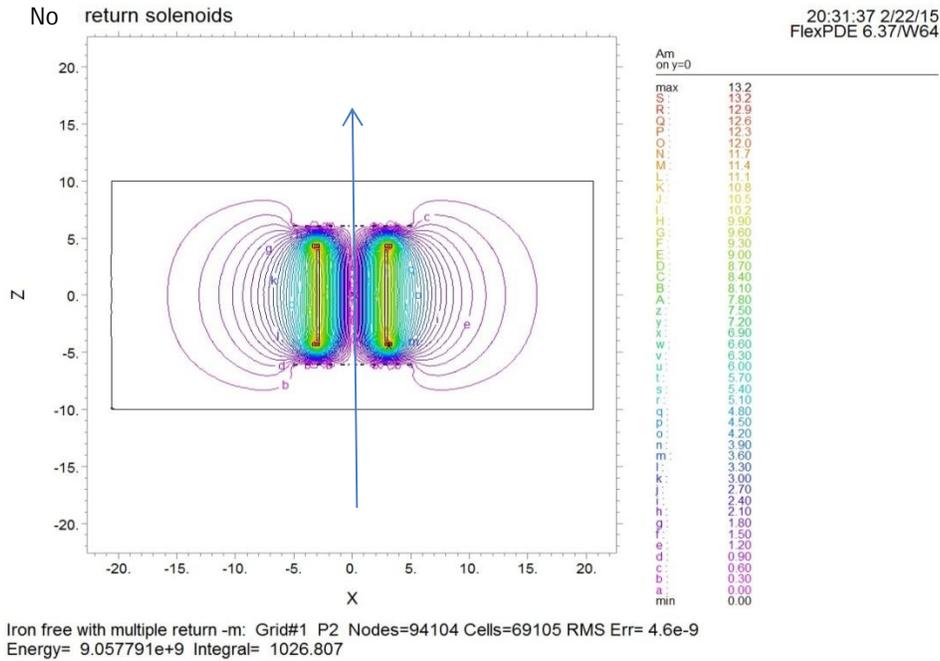

**Figure 6.** Lines of magnetic field in detector. Detector axis is running in a vertical direction (marked by arrow). One can see how the end coils divert the flux from longitudinal direction. X and Z measured in meters.

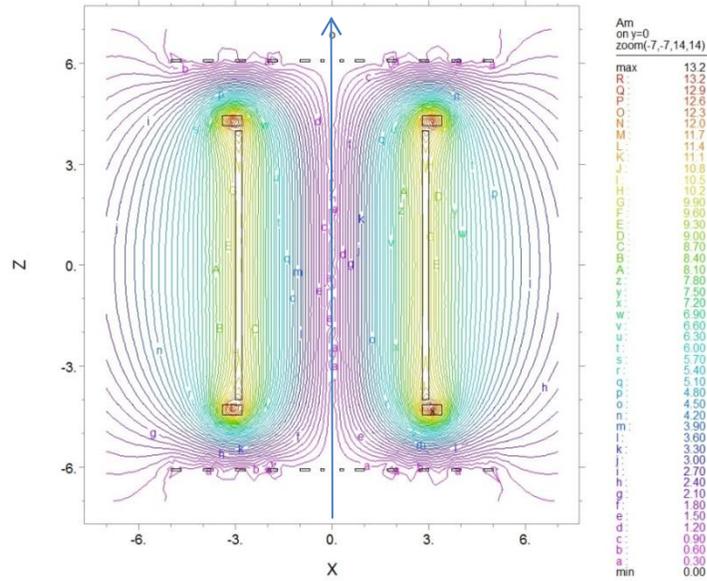

**Figure 7.** Lines of magnetic field from previous Fig. 6 which is zoomed here. X and Z measured in meters.

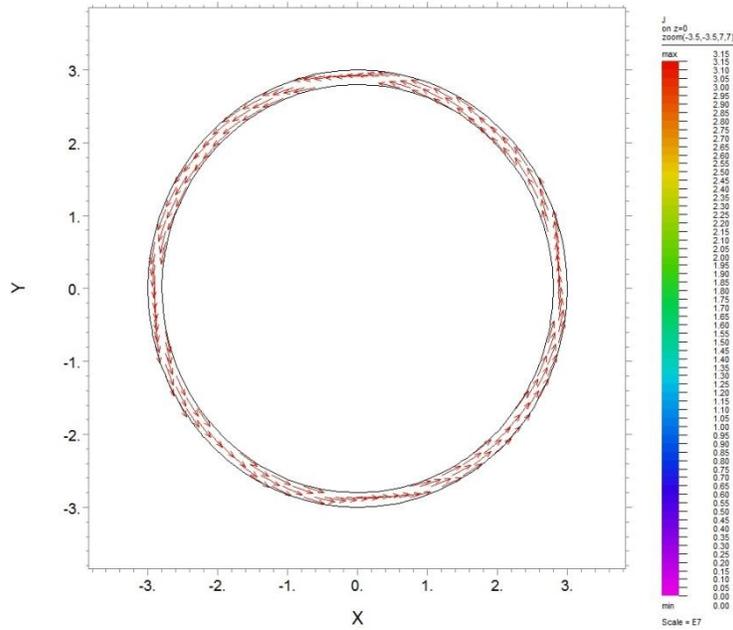

**Figure 8.** Distribution of currents in a central solenoid shown by vectors, top view. X and Y measured in meters.

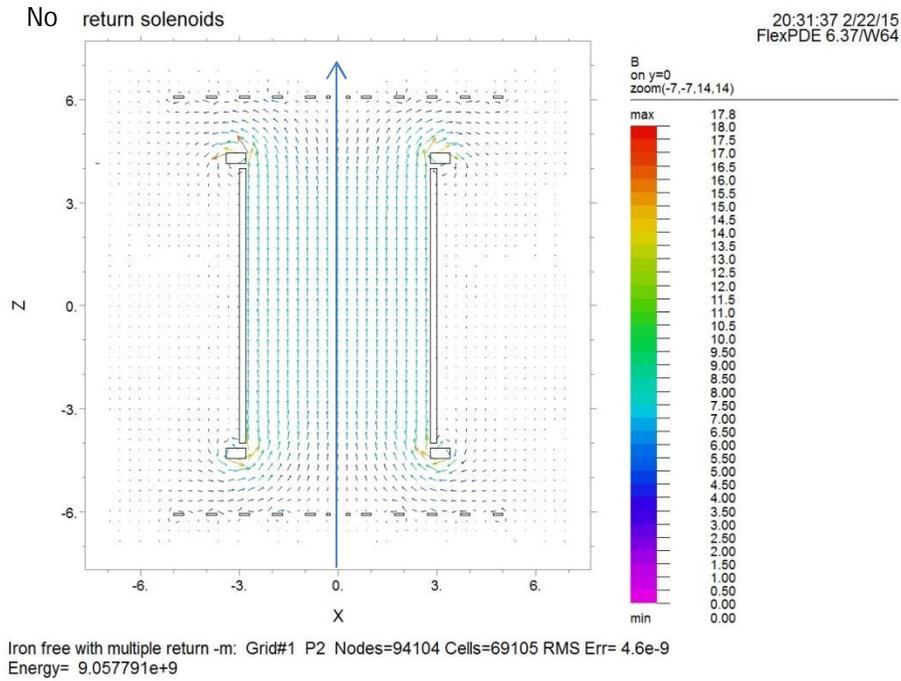

**Figure 9.** Vectors of magnetic field. X and Z measured in meters.

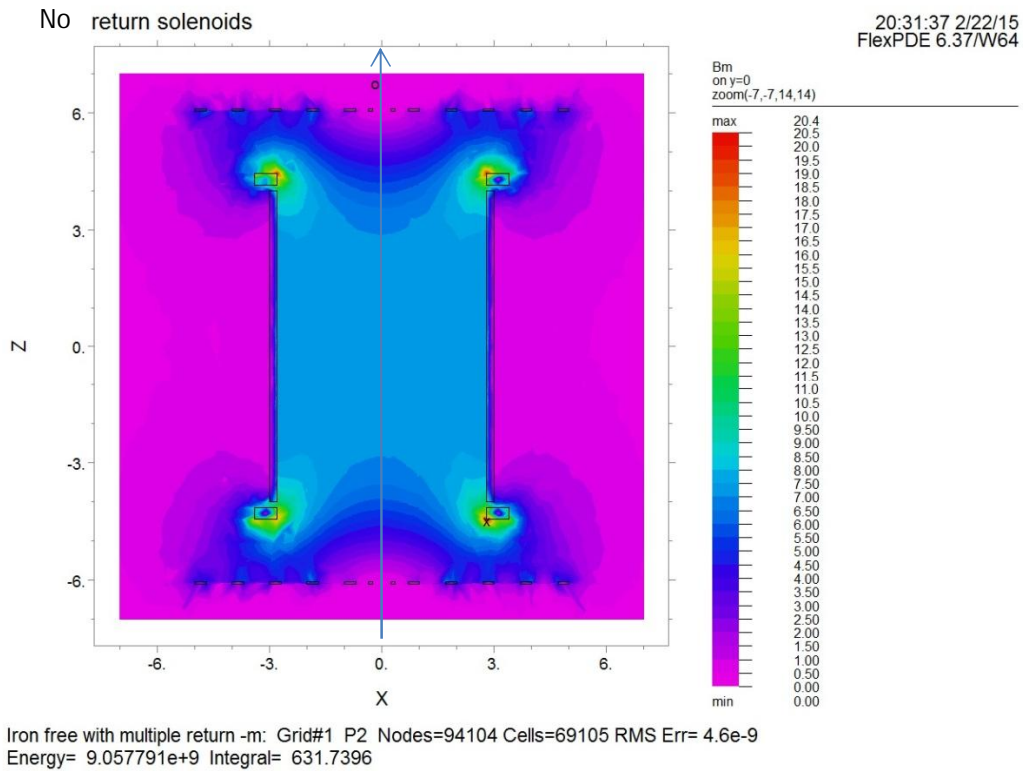

**Figure 10.** Magnetic field amplitude contour, painted. One can see, that is longitudinal direction ±3*m*, the field inside solenoid is pretty homogenous. Coordinates measured in meters.

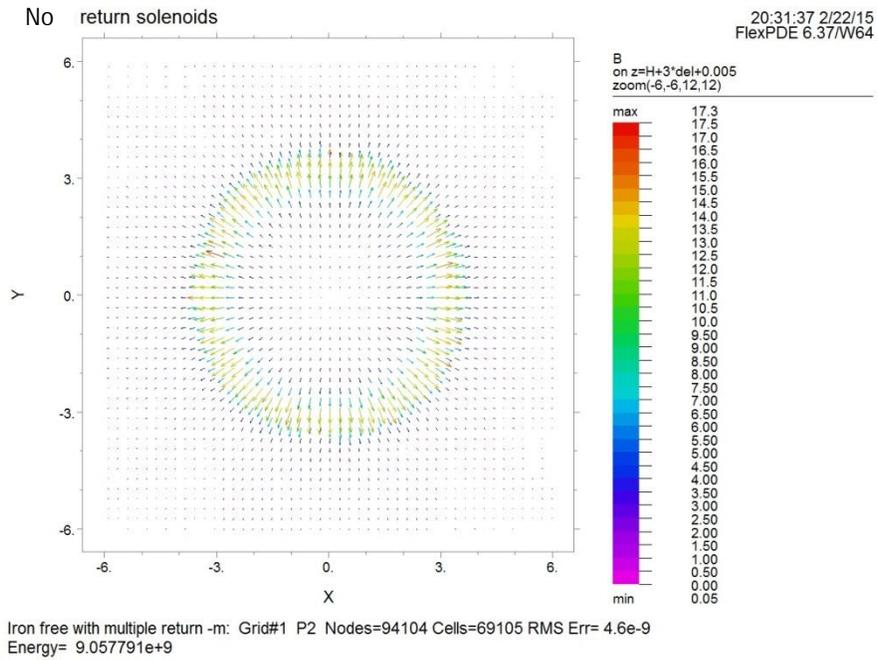

**Figure 11.** Shown are the vectors of magnetic field in a plane just above cutoff of central solenoid. X and Y are measured in meters.

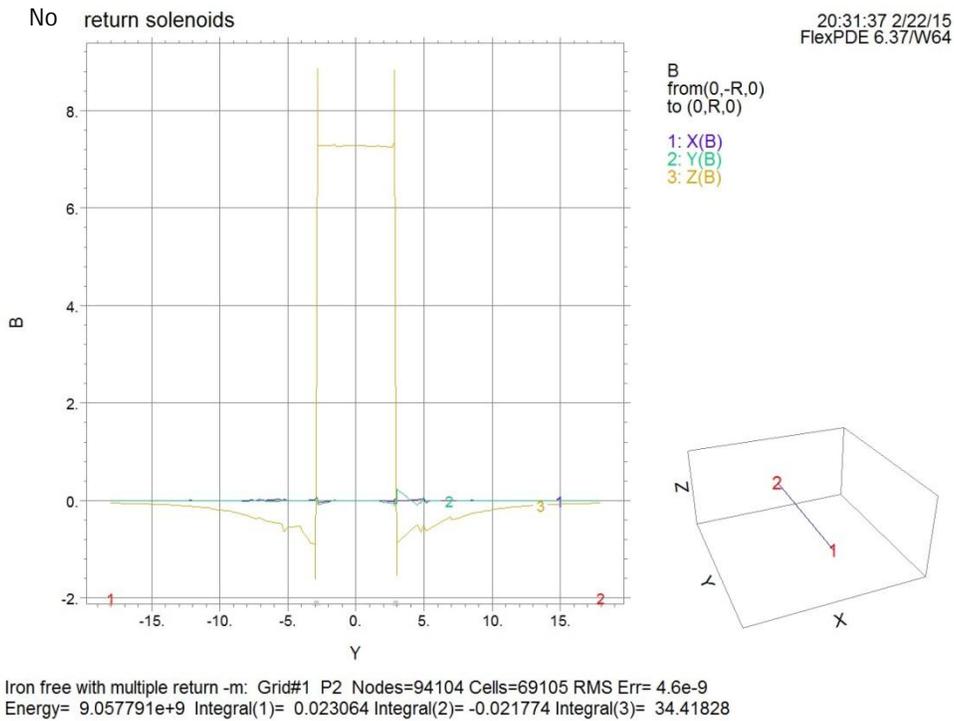

**Fgure 12**. Elevation of magnetic field across system in a midplane. Y-coordinate in meters, B in Tesla.

From Figure 12 one can see, that the field level at outer region off solenoid has maximum ~1$T$ and drops to ~200 $G$ at distance ~20 $m$.

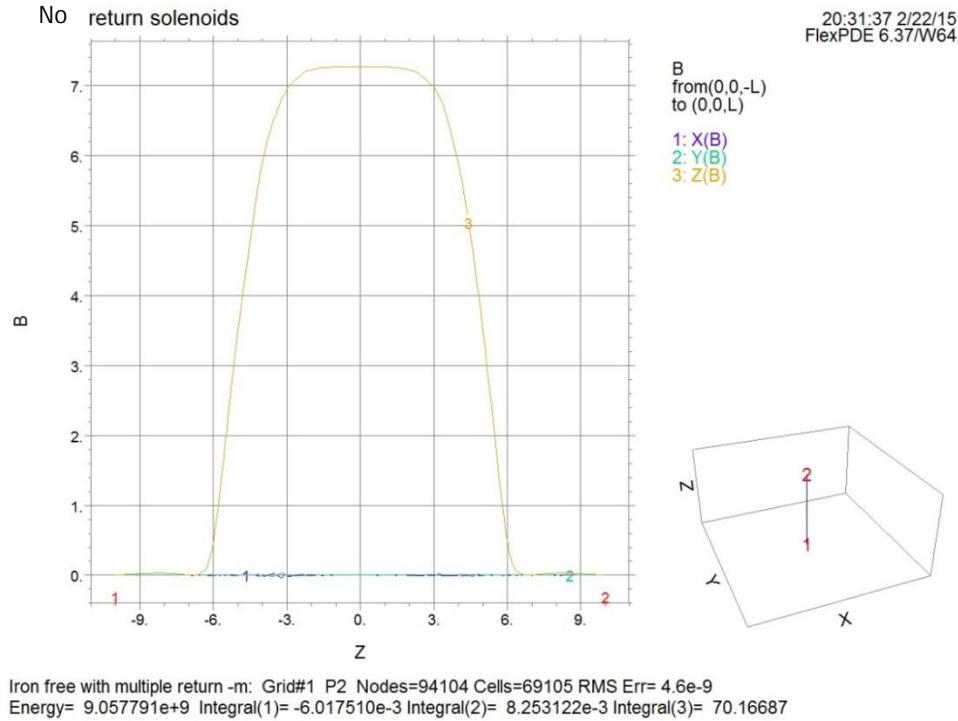

**Figure 13**. Elevation of magnetic field along central axis. Field value in *Tesla*.

We would like to underline, that with multiple-return solenoids, the field at max will be $8T$ for the same currents in central solenoids as they are in Fgs.12, 13. So the difference in not big, however; this we explained by the fact that the flux now becomes redistributed in unlimited radial space, so the input of outer field into integral $\oint \vec{B} d\vec{l} = \mu_0 I_{tot}$ is not big. Remember, that the field outside of long solenoid is zero.

## CONCLUSIONS

Single solenoidal system demonstrates acceptable field profile. In a longitudinal direction it behaves as usual Iron-free system with very restricted field propagation along the beam trajectory. In transverse direction the field drop is not so fast if no coils at outer region are present, but this is irrelevant. Mostly undesirable is propagation of fields in longitudinal direction as the leaking fields could influence on the beam trajectory. In transverse direction these field could produce a discomfort to neighboring second detector, which is planned to be used in push-pull mode. The field level at 20 m from axis drops to ~200*Gauss* which could be compensated easily.

Mentioned system of room temperature coils in our publication [2] at the outer radius could restrict propagation of field in transverse direction. This means also that outer solenoid in dual or triple solenoidal systems could be made as a few separated axial coils separated longitudinally. At some optimization these coils could also be room-temperature ones. The concentric end-cap coils are eliminating propagation of field in a longitudinal direction already; they are running at room-temperature.

The disadvantage of a single-solenoidal system is that the return field indeed becomes a small-level one which makes muon spectrometry less effective.

Absence of saturated thick and massive Iron yoke makes possible reversing field in a detector faster; this might be useful for exclusion of asymmetries of registration system. Ramp rate from 1 to 3 *kA/s* looks feasible [4]. The only restriction for the rate of switching for the system of coils embedded in Al carcasses (one version of the coil design) defined by induced currents, but this could be tolerable. Other limiting factor might be associated with PS. Just reminding, that we have suggested 5-6 independent PSs for the central multi-layer solenoid and about the same number of PSs for the coils at the ends of central solenoids see Fig.4. The end coils are also feed by separate PSs. The option with the continuous conductor for the end cap coils required manipulation with the linear current density along radial direction, so it is less attractive.